\documentclass[12pt,preprint]{aastex}
\begin{document}
%\tighten
\def\etal{{\it et al.\/}}
\def\cf{{\it cf.\/}}
\def\ie{{\it i.e.\/}}
\def\eg{{\it e.g.\/}}

\title{On particle acceleration around shocks II: a fully general
method for arbitrary shock velocities and scattering media}
\author{{\bf Pasquale Blasi$^1$ and \bf Mario Vietri$^2$}}
\affil{$^1$ INAF/Osservatorio Astrofisico di Arcetri, Firenze, Italy\\
$^2$ Scuola Normale Superiore, Pisa, Italy}
{}
\begin{abstract}
The probability that a particle, crossing the shock along a given
direction, be reflected backwards along another direction, was
shown to be the key element in determining the spectrum of
non--thermal particles accelerated via the Fermi mechanism around a
plane--parallel shock in the test--particle limit. Here an
explicit equation for this probability distribution is given, for
both the upstream and downstream sections. Though analytically
intractable, this equation is solved numerically, allowing the
determination of the spectrum in full generality, without
limitation to shock speed or scattering properties. A number of
cases is then computed, making contact with previous numerical
work, in all regimes: Newtonian, trans--relativistic, and fully
relativistic.
\end{abstract}


\section{Introduction}

It is possible to formulate particle acceleration around shocks,
in the test particle limit, in an exact form, for arbitrary shock
speeds, and arbitrary anisotropies in the particle distribution
function (Paper I, Vietri 2003). It turns out that the physically
relevant boundary conditions are those at the shock, not those at
downstream infinity. These conditions at the shock fix the
particle spectrum which was shown to be (under some simplifying
assumptions) an {\bf exact} power--law in the particle momentum,
independent of the shock speed and of all assumptions about the
scattering properties of the medium.

The exact value of the spectral index depends in a detailed way
upon the diffusion properties of the media involved. It was shown
in Paper I that this dependence is encapsulated in two functions,
$P_u(\mu_\circ,\mu)$ and $P_d(\mu_\circ,\mu)$ respectively, which
denote the conditional probabilities that a particle entering the
upstream (downstream) medium along a direction $\mu_\circ$, will
leave it for the downstream (upstream) section along a direction
$\mu$\footnote{In these expressions, all angles are measured in
the fluid frame.}. In paper I (but see also Freiling, Vietri and
Yurko 2003 for some tricky mathematical details), these functions
were built from the eigenfunctions of the angular part of the
scattering problem. Though conceptually satisfactory, this
procedure is however computationally tricky, and one may wonder
whether, given the scattering law for the particles, it may be
possible to determine an equation for the scattering probabilities
that yields them directly, without having to build the infinitely
many eigenfunctions of the angular equation. It turns out that
this is actually possible, by drawing upon the analogy with
scattering atmospheres (Chandrasekhar 1949), and by using an
obvious symmetry property of the problem at hand.

It is the purpose of this paper to
present this derivation, and a handful of illustrative
computations, with the major aim of showing the ability of
the method to reproduce well--established results, drawn
from the literature. A more detailed analysis of the spectra
in several interesting cases, including the important
hyperrelativistic regime, will be presented elsewhere.

The plan of the paper is as follows: in section 2, we present the
derivation of the equations which determine $P_d(\mu_\circ,\mu)$
and $P_u(\mu_\circ, \mu)$. In section 3, we show that a diffusive
equation identical to that originally given by Kirk and Schneider
(1987) is recovered in the Fokker--Planck regime (or
small--pitch--angle scattering, SPAS) of our scattering equation,
Eq. \ref{main} . In section 4, we describe our full method to compute
numerically the spectral slope and angular distribution function.
In section 5 we show that our method reproduces very well results
which have already appeared in the literature, in the newtonian,
trans-relativistic and fully relativistic regimes (the ultra-relativistic
regime will be discussed in forthcoming publication). We conclude
in section 6.

\section{The equations for $P_d$ and $P_u$}

In paper I, we showed that a suitable, relativistically covariant
(though not manifestly so) equation for the transport of the
particle distribution function for arbitrary pitch angle scattering
in the presence of a large-coherence length magnetic field is:
\begin{equation}
\gamma (u+\mu) \frac{\partial f}{\partial z} = \int\left(
-w_e(\mu',\mu,\phi,\phi')
f(\mu,\phi)+w_e(\mu,\mu',\phi,\phi')f(\mu',\phi')\right)
d\!\mu'd\!\phi' + \omega\frac{\partial f}{\partial\phi}\;.
\end{equation}
In this equation, $z$ is the distance from the shock along the shock
normal, in the shock frame. All other quantities are measured in the
fluid frame: $u$ is the fluid speed with respect to the shock, in
units of $c$, and $\gamma =1/\sqrt{1-u^2}$ its associated Lorentz
factor, $f$ is the DF of particles with impulse $p$ in the fluid
frame (which, by assumption, does not change because we assumed
pitch--angle scattering), and $w_e(\mu,\mu',\phi,\phi')$ is the
scattering probability that a particle, originally moving along a
direction making an angle with the shock normal such that its cosine
is $\mu'$, and along $\phi'$, be deflected along a new direction
$\mu,\phi$. The fluid is assumed to move in the positive $z$
direction, so that its speed is at all places $>0$, and the angles
are measured from the direction of motion of the fluid.

The scattering is assumed linear (thus these results apply only in
the test particle approximation), but the scattering angle is
nowhere assumed small: we shall derive results for the diffusive
approximation in a later section. This equation is the
plane--geometry, fully relativistic version of an equation
originally developed by Gleeson and Axford (1967) for spherically
symmetric, non--relativistic shocks.

Compared with Eq. 7 of Paper I, the injection has been dropped
because we are considering particles at very large energies, much
larger than the injection energy; we now also specialize to the case
in which no long-coherence length magnetic field is present. The
treatment of the problem in full generality will be presented
elsewhere (Morlino, Blasi and Vietri, in preparation).

Besides dropping the term $\omega(\partial f/\partial\phi)$, we
remark that the system has now become symmetric for rotations around
the shock normal, so that the DF $f$ will be independent of the
angle $\phi$. Furthermore, the elemental scattering probability
$w_e$ can depend only on the angle $\phi-\phi'$ between the incoming
and outgoing direction, $w_e = w_e(\mu,\mu',\phi-\phi')$, even
though it can depend separately upon $\mu$ and $\mu'$. In this
symmetric case, we can average the previous equation over either
$\phi$ or $\phi'$, obtaining
\begin{equation}{}
\label{main}\label{eq:transport} \gamma (u+\mu) \frac{\partial
f}{\partial z} = \int\left( -w(\mu',\mu)
f(\mu)+w(\mu,\mu')f(\mu')\right) d\!\mu'd\!\phi'\;,
\end{equation}
where
\begin{equation}\label{averagew}
w(\mu,\mu') \equiv \frac{1}{2\pi}\int w_e(\mu,\mu',\phi-\phi')
d\!\phi\;,
\end{equation}
an equation which will be used later on.

\subsection{The downstream case}

In order to fix the ideas, we consider the problem of determining
$P_d$, the conditional probability for the downstream frame; we
shall explain later how to adapt these results to the upstream
frame. Take the origin of coordinates to be located at the shock,
the downstream section to be at $z>0$, and define $u$ as the {\bf
modulus} of the shock speed with respect to the fluid. We are using
variables in the fluid frame: $\mu$ is thus the cosine of the angle
between the particle speed and the shock normal, {\it in the fluid
frame}. This means that the particles which can cross from upstream
to downstream (\ie, those with a positive component along the shock
normal of the velocity {\it with respect to the shock}) are those
with
\begin{equation}
u+\mu > 0\;,
\end{equation}
while those that cross the shock backwards, from the downstream to
the upstream section, have
\begin{equation}
u+\mu < 0\;.
\end{equation}

Now let $f$ represent a beam of particles, all entering the
downstream section along the same direction, $\mu_\circ$; we showed
in paper I that the particle flux across a surface fixed in the
shock frame, but expressed in downstream coordinates, is given by:
\begin{equation}
F \equiv F_\circ\gamma = \int \gamma (u+\mu) f d\!\mu\;.
\label{fluxdef}
\end{equation}
We consider now a pencil beam of particles, all moving initially
in the direction $\mu_\circ$. Thus, at the shock:
\begin{equation}
\label{fattheboundary1}
f(\mu) = \frac{F_\circ}{u+\mu}\delta(\mu-\mu_\circ) \;\;\;\;\;\;\; u+\mu
\geq 0\;,
\end{equation}
with $F_\circ$ an obvious normalization. It was shown in paper I
that Eq. \ref{fattheboundary1} provides a suitable boundary
condition for Eq. \ref{main}, which then returns the outgoing
flux; by definition, this is precisely the required $P_d$. We thus
have
\begin{equation}\label{fattheboundary2}
-(u_d+\mu) f(\mu) = F_\circ P_d(\mu_\circ,\mu)\;\;\;\;\;\; u+\mu < 0\;.
\end{equation}

To begin, let us define the inward ($f_+ \equiv f(\mu) |_{u+\mu> 0}$) and
outward ($f_- \equiv f(\mu)|_{u+\mu<0}$) parts of the distribution
function, and let us consider a surface at a distance $z$ from the shock,
fixed in the shock frame. The flux of particles moving backwards toward
the shock, through this surface, is:
\begin{equation}
\gamma (u+\mu) f_-(\mu) = - \int_{-u}^1 f_+(\mu') \gamma (u+\mu')
P_d(\mu',\mu) d\!\mu'\;.
\end{equation}
The minus sign accounts for the fact that $u+\mu < 0$, while all terms
inside the integral are positive.

Strictly speaking, in the above equation, we should have written
$P_d(\mu',\mu,z)$ instead of simply $P_d(\mu',\mu)$. In other
words, for a general situation, we cannot assume that the
conditional probability $P_d$ be independent of its location
inside the downstream region. However, in this case this
assumption is fully justified, because the downstream region is
semi--infinite: in other words, it remains identical to itself
whenever we add or subtract finite regions, and its diffusive
properties also remain unaltered whenever we add or subtract
finite regions. It follows that the conditional probability $P_d$
from some finite $z$ to downstream infinity must be exactly
identical to that from $0$ to downstream infinity. This invariance
principle, and this whole discussion, are identical to those in
Chandrasekhar (1949, Chapter 4, Sections 28-29), and provide an
exact justification for the equation above.

We now differentiate the above with respect to $z$:
\begin{equation}\label{star}
\gamma(u+\mu)\frac{\partial f_-}{\partial z} = -\int_{-u}^1 d\!\mu'
\gamma(u+\mu')\frac{\partial f_+}{\partial z} P_d(\mu',\mu)
\end{equation}
while Eq. \ref{main} gives us
\begin{equation}\label{mainminus}
\gamma(u+\mu) \frac{\partial f_-}{\partial z} = -d(\mu) f_- +
\int_{-u}^1 d\!\mu' w(\mu,\mu') f_+(\mu') +
\int_{-1}^{-u}d\!\mu' w(\mu,\mu') f_-(\mu')
\end{equation}
\begin{equation}\label{mainplus}
\gamma(u+\mu) \frac{\partial f_+}{\partial z} = -d(\mu) f_+ +
\int_{-u}^1 d\!\mu' w(\mu,\mu') f_+(\mu') +
\int_{-1}^{-u}d\!\mu' w(\mu,\mu') f_-(\mu')\;,
\end{equation}
where we used the shorthand
\begin{equation}
d(\mu) \equiv \int w(\mu',\mu) d\!\mu'\;.
\end{equation}
We now evaluate Eq. \ref{mainminus} at $z = 0$, using Eqs. \ref{fattheboundary1} and
\ref{fattheboundary2}, to obtain
\begin{equation}\label{inter1}
\gamma(u+\mu) \frac{\partial f_-}{\partial z}|_\circ =
\frac{F_\circ d(\mu)}{\gamma(u+\mu)} P_d(\mu_\circ, \mu) +
\frac{F_\circ w(\mu,\mu_\circ)}{\gamma(u+\mu_\circ)}-
\int_{-1}^{-u}d\!\mu' \frac{w(\mu, \mu') F_\circ
P_d(\mu_\circ,\mu')}{\gamma(u+\mu')} \;.
\end{equation}
We can now proceed analogously for Eq. \ref{mainplus}: we again evaluate it at $z = 0$,
and use Eqs. \ref{fattheboundary1} and \ref{fattheboundary2}, to obtain:
\begin{equation}\label{inter2}
\gamma(u+\mu) \frac{\partial f_+}{\partial z}|_\circ = -
\frac{F_\circ d(\mu)}{\gamma(u+\mu)} \delta(\mu-\mu_\circ) +
\frac{F_\circ w(\mu,\mu_\circ)}{\gamma(u+\mu_\circ)}-
\int_{-1}^{-u}d\!\mu' \frac{w(\mu, \mu') F_\circ
P_d(\mu_\circ,\mu')}{\gamma(u+\mu')} \;.
\end{equation}
As our last step, we plug Eqs. \ref{inter1} and \ref{inter2} into Eq. \ref{star} to obtain:
\begin{eqnarray}\label{mainPd}
P_d(\mu_\circ,\mu)
\left(\frac{d(\mu_\circ)}{u+\mu_\circ}-\frac{d(\mu)}{u+\mu}
\right) = \frac{w(\mu,\mu_\circ)}{u+\mu_\circ} + \int_{-u}^1
d\!\mu' \frac{P_d(\mu',\mu) w(\mu',\mu_\circ)}{u+\mu_\circ} -
\nonumber \\
 \int_{-1}^{-u} d\!\mu' \frac{P_d(\mu_\circ,\mu')
w(\mu,\mu')}{u+\mu'}  - \int_{-u}^1 d\!\mu' P_d(\mu',\mu)
\int_{-1}^{-u} d\!\mu'' \frac{w(\mu',\mu'')
P_d(\mu_\circ,\mu'')}{u+\mu''}
\label{eq:Pd}
\end{eqnarray}
which is the sought--after result: an equation relating $P_d$ only
to the diffusion probability $w$.

\subsection{The upstream case}

A completely analogous derivation holds in this case. The pencil beam
is now entering the upstream section, so that, at the shock, we have:
\begin{equation}\label{fattheboundary3}
f(\mu) = -\frac{F_\circ}{u+\mu}\delta(\mu-\mu_\circ)\;\;\;\;\;\; u+\mu \leq 0
\end{equation}
and we have the obvious definition, from paper I:
\begin{equation}\label{fattheboundary4}
f(\mu) = \frac{F_\circ}{u+\mu} P_u(\mu_\circ,\mu)\;\;\;\;\;\; u+\mu
> 0\;.
\end{equation}
At a distance $z$ from the shock the flux of particles moving
toward the shock is
\begin{equation}\label{star2}
\gamma(u+\mu)\frac{\partial f_+}{\partial z} = -\int_{-1}^{-u}
d\!\mu' \gamma(u+\mu') P_u(\mu',\mu) \frac{\partial f_-}{\partial
z}
\end{equation}
where the same argument about the use of $P_u(\mu',\mu)$ instead
of $P_u(\mu',\mu,z)$ applies as above. We may evaluate the above
equation at $z=0$ using Eqs. \ref{mainminus} and \ref{mainplus}
(which still hold), and Eqs. \ref{fattheboundary3} and
\ref{fattheboundary4}, and plug the result into Eq. \ref{star2},
to obtain
\begin{eqnarray}\label{mainPu}
P_u(\mu_\circ,\mu)\left(\frac{d(\mu_\circ)}{u+\mu_\circ}-\frac{d(\mu)}{u+\mu}
\right)= \frac{w(\mu,\mu_\circ)}{u+\mu_\circ} - \int_{-u}^1
d\!\mu' \frac{w(\mu,\mu')P_u(\mu_\circ,\mu')}{u+\mu'}+ \nonumber\\
\int_{-1}^{-u}d\!\mu'\frac{w(\mu',\mu_\circ)P_u(\mu',\mu)}{u+\mu_\circ}
-\int_{-1}^{-u} d\!\mu' P_u(\mu',\mu)\int_{-u}^1
d\!\mu''\frac{w(\mu',\mu'') P_u(\mu_\circ,\mu'')}{u+\mu''}
\label{eq:Pu}
\end{eqnarray}
which is our other final result.
{}

%\section{The Newtonian limit}

\section{The small--pitch--angle limit}\label{sec:fp}

In this section, we wish to derive the small pitch--angle
scattering (SPAS) limit of our scattering equation, which
holds for arbitrary deflection angles, not just small ones.
Before proceeding with the derivation of the Fokker--Planck
equation, we need to establish a very useful result. The function
$w$ is not completely arbitrary, but is subject to the following
requirement. We expect that, when the distribution function (DF)
is isotropic ({\it i.e.}, when $f(\mu) = \mbox{const.}$), there
will be no further evolution: in other words, we expect in this
case the RHS of the above equation to vanish. For this to occur,
we shall not take the obvious condition
\begin{equation}
\int_{-1}^{+1} w(\mu,\mu') d\!\mu' =
\int_{-1}^{+1} w(\mu',\mu) d\!\mu'
\end{equation}
but its stronger version,
\begin{equation}
w(\mu,\mu') = w(\mu',\mu)\;.
\end{equation}
The above equation is a statement of detailed balance, which we
may assume to hold because these scattering processes are exactly
those which lead to isotropization of the DF, if thermal
equilibrium were to hold. In our case we have neglected (for
excellent, well--known reasons) the exchange of energy of the
particles with the background fluid, so we cannot expect true
thermal equilibrium to hold, but since we have included all major
scattering processes, we expect equilibrium ({\it i.e.}, detailed
balance except for energy equipartition) to hold at least in so
far as the angular part is concerned. It is probably worthwhile to
remark that a similar assumption on $w$ is made by Chandrasekhar
(1949, Ch. IV, Section 31, Eq. 29-1).

We now rewrite Eq. \ref{main} by means of a new scattering
probability, $W(\mu';\alpha)$, to be defined as follows:
\begin{equation}
W(\mu';\alpha) d\!\alpha \equiv w(\mu'+\alpha, \mu')
d\!(\mu'+\alpha)\;.
\end{equation}
$W$ is simply the probability that a particle is deflected from
its initial direction along $\mu'$, to a new direction
$\mu'+\alpha$. By means of this new quantity Eq. \ref{main} can be
recast as:
\begin{equation}
\label{main2}
\gamma (u+\mu) \frac{\partial f}{\partial z} =
\int_{-1}^{+1} d\!\alpha \left( W(\mu-\alpha;\alpha) f(\mu-\alpha) -
W(\mu;-\alpha)f(\mu)\right)
\end{equation}
while the detailed balance equation becomes:
\begin{equation}
W(\mu-\alpha;\alpha) = W(\mu; -\alpha)\;.
\end{equation}

The scattering equation in the form \ref{main2} is exactly
identical to that given by Landau and Lifshtiz (1984, Section 21,
Eq. 21.1), so that their derivation of the Fokker--Planck
immediately applies. Please note however that our sign convention
differs from theirs: what we call $\alpha$ is for them $-q$.
Following step by step their derivation we find that
\begin{equation}
\gamma(u+\mu)\frac{\partial f}{\partial z} \approx \frac{\partial}{\partial \mu}
\left(A f + B\frac{\partial f}{\partial \mu}
\right)\;,
\end{equation}
with the definitions
\begin{equation}
\label{coeff1} A \equiv  -\int_{-\infty}^{+\infty} d\!\alpha\;
\alpha W(\mu;\alpha) + \frac{\partial B}{\partial \mu}
\end{equation}
\begin{equation}\label{coeff2}
B \equiv  \frac{1}{2} \int_{-\infty}^{+\infty} d\!\alpha
\;\alpha^2 W(\mu;\alpha)\;.
\end{equation}
In the two integrals above, we have taken the integration range to
extend from $-\infty$ to $+\infty$, because we have assumed the
validity of the SPAS regime, in which case $W(\mu;\alpha)$ has one
very strong, narrow peak around $\alpha = 0$:

We show now that, as a consequence of detailed balance, $A = 0$. We first
notice that
\begin{equation}
\frac{\partial}{\partial \mu} W(\mu;\alpha) =
\frac{\partial}{\partial \mu} W(\mu'; \mu-\mu')
\end{equation}
where $\mu' = \mu+\alpha$, because of detailed balance, so that
\begin{equation}
\frac{\partial}{\partial \mu} W(\mu;\alpha) =
\frac{\partial}{\partial \alpha} W(\mu';\alpha)\;.
\end{equation}
We use the above into
\begin{equation}
\frac{\partial B}{\partial \mu} =
\frac{1}{2}\int_{-\infty}^{+\infty} d\!\alpha \;\alpha^2
\frac{\partial}{\partial \mu} W(\mu;\alpha) =
\frac{1}{2}\int_{-\infty}^{+\infty} d\!\alpha \;\alpha^2
\frac{\partial}{\partial \alpha} W(\mu'; \alpha)
\end{equation}
which can now be integrated by parts to yield
\begin{equation}
\frac{\partial B}{\partial \mu} = -\int_{-\infty}^{+\infty}
d\!\alpha\; \alpha W(\mu'; \alpha)\;.
\end{equation}
We now use again detailed balance, and the variable change $x \equiv -\alpha$
in the definite integral above, to obtain
\begin{equation}
\frac{\partial B}{\partial \mu} = \int_{-\infty}^{+\infty}
d\!\alpha \;\alpha W(\mu; \alpha),
\end{equation}
which, when inserted into Eq. \ref{coeff1}, yields $A = 0$. We
thus obtain, for the scattering equation in the SPAS, or
Fokker--Planck, limit:
\begin{equation}\label{final}
\gamma(u+\mu)\frac{\partial f}{\partial z} = \frac{\partial}{\partial \mu}
\left(B \frac{\partial f}{\partial \mu}\right)\;,
\end{equation}
with $B$ given by Eq. \ref{coeff2}. The use of detailed balance
implies that the scattering integral reduces, in the SPAS limit,
to the divergence of a vector proportional to the gradient of the
DF, exactly like in the classical heat conduction problem.

As a simple test of this treatment, we now re-derive a well--known
result, {\it i.e.}, that, in the isotropic limit,
\begin{equation}\label{iso}
B = K (1-\mu^2)
\end{equation}
with $K$ a constant. The isotropic limit means that the scattering
coefficient probability $W$ can only depend on the angle $\theta-\theta'$
between the directions of motion. We shall take thus
\begin{equation}
W(\mu;\alpha) = N f(\xi)
\end{equation}
where
\begin{equation}
\xi \equiv  \frac{\sin(\theta-\theta')}{\sigma}\;,
\end{equation}
with $\sigma$ a parameter which we take $\sigma\ll 1$, in order to
enforce the SPAS regime. Also, we assume that $f(x)$ is an even
function of $x$, with a single minimum around $x = 0$ (again, SPAS
regime); we also assume, for the moment, that $W$ is normalized to
unity:
\begin{equation}\label{norm}
\int_{-\infty}^{+\infty} d\!\alpha N f(\alpha) = 1\;,
\end{equation}
and will come back to this assumption in a moment.

We now compute $B$ from Eq. \ref{coeff2}. Since we shall be taking
the limit $\sigma\rightarrow 0$, only the argument
$sin(\theta-\theta') \approx \sigma \ll 1$ will contribute to the
integral, so that, defining $\epsilon \equiv \theta - \theta'$, we
find
\begin{equation}
\xi \approx \frac{\epsilon}{\sigma}
\end{equation}
\begin{equation}
\alpha \equiv \mu'-\mu = \cos\theta' - \cos\theta = \cos(\theta+\epsilon)-
\cos\theta \approx -\sin\theta\epsilon
\end{equation}
so that the normalization condition, Eq. \ref{norm} requires
\begin{equation}
N = \left(\sin\theta \sigma\int_{-\infty}^{+\infty}
f(\frac{\epsilon}{\sigma}) d\!\frac{\epsilon}{\sigma}
\right)^{-1}\;.
\end{equation}
$B$ then becomes
\begin{eqnarray}
B = \frac{1}{2}\int_{-\infty}^{+\infty} d\!\alpha \;\alpha^2
W(\mu;\alpha) \approx \frac{1}{2} N
\sigma^3\sin^3\theta
\int_{-\infty}^{+\infty} f(\xi) \xi^2
d\!\xi = \nonumber\\
\sin^2\theta\sigma^2
\frac{\int_{-\infty}^{+\infty} \xi^2 f(\xi)
d\!\xi}{\int_{-\infty}^{+\infty} f(x) d\!x}
\end{eqnarray}
which is clearly identical to Eq. \ref{iso}. If we now assume $W$
to have arbitrary normalization (because it represents the
probability of scattering per unit time), we see that the above
derivation is changed only through an inessential multiplicative
constant. This completes our derivation.

\subsection{Scattering probability}

Later on, we shall compare numerical results based upon our new
approach with those which previous authors have obtained in the
fully isotropic, small pitch angle scattering limit. In order to do
this, we have to determine a suitable form for $w$ in this case. It
is clear that any function very strongly peaked around the forward
direction simulates the SPAS regime, like, for instance,
\begin{equation}
w \propto e^{-(\mu-\mu')^2/2\sigma^2}\;,\;\;\;\; \sigma \ll 1\;.
\end{equation}
However, not all such peaked functions reproduce the isotropic case,
which is defined as that in which the {\it elemental} scattering
probability $w_e$ is independent of anything except for the
deflection angle:
\begin{equation}
w_e(\mu,\mu',\phi,\phi') = w(\cos\Theta)
\end{equation}
with $\Theta$ given by
\begin{equation}
\cos\Theta \equiv \mu\mu' +
\sqrt{1-\mu^2}\sqrt{1-\mu'^2}\cos(\phi-\phi')\;.
\end{equation}
We have seen above that the transport equation in the SPAS regime
does not depend upon the exact form of the scattering function
$w_e$, so that we are free to choose any analytically tractable one.
We choose:
\begin{equation}
w_e(\cos\Theta) =
\frac{1}{\sigma}e^{-\frac{1-\cos\Theta}{\sigma}}\;,\;\;\;\; \sigma
\ll 1\;,
\end{equation}
which is obviously suitably normalized to unity when the integral is
made over all scattering angles. Using Eq. \ref{averagew} we find
now
\begin{equation}\label{wpeaked}
w = \frac{1}{2\pi}\int_0^{2\pi}w_e d(\phi-\phi') =
\frac{1}{\sigma}e^{-\frac{1-\mu\mu'}{\sigma}}
I_0(\frac{\sqrt{1-\mu^2}\sqrt{1-\mu'^2}}{\sigma})
\end{equation}
where $I_0(x)$ is Bessel's modified function of order $0$, and use
has been made of the formula (Gradshteyn and Ryzhik, 1994, 3.339):
\begin{equation}
\int_0^\pi \exp(z\cos x) d\!x = \pi I_0(z)\;.
\end{equation}
In the following we shall use Eq. \ref{wpeaked} when computing
spectra, taking $\sigma \ll 1$ for the isotropic SPAS limit and
$\sigma\gg 1$ for the isotropic large angle scattering regime.

\section{Calculation of the spectrum and angular distribution function}
\label{sec:numer}

In this section we outline the procedure to follow in order to calculate the
spectral slope and the angular distribution function of the accelerated
particles, for an arbitrary choice of the shock velocity and of the
scattering properties of the fluid.

For the purpose of making contact with previous literature on the
subject, we carry out our calculations with a scattering function
$w(\mu,\mu')$ as given in Eq. \ref{wpeaked}, choosing a small value
of the parameter $\sigma$ for the case of small pitch angle
scattering, and a large value of $\sigma$ to reproduce the regime of
large angle scattering. Eq. \ref{wpeaked} ensures that the
scattering remains isotropic in both regimes. However, we wish to
stress that we elected to use an isotropic scattering in order to
make contact with all other authors (so far), who have only used
isotropic scattering, even though this is by no means required by
our method, which can effectively deal with anisotropic scattering.

It is worth stressing that our method does not require any intrinsic
approximations or assumptions on the type of scattering; the only
approximations are due to round-off in the numerical solution of the
integral equations for $P_d$ and $P_u$.

The procedure for the calculation of the slope of the spectrum of
accelerated particles is as follows: for a given Lorentz factor of
the shock ($\gamma_{sh}$), the velocity of the upstream fluid
$u=\beta_{sh}$ is calculated. We then determine the velocity of
the downstream fluid $u_d$ from the conventional jump conditions
at the shock, after assuming some equation of state for the
post--shock fluid. Given $u$, $u_d$ and a scattering function $w$,
we can solve the integral equations for $P_u$ and $P_d$
iteratively. We recall here that these are functions of an
entrance and an exit angle, and the convergence of the method can
be numerically time consuming, in particular for small values of
$\sigma$. Aside from this numerical issue, one should also
remember that our equations for $P_u$ and $P_d$ are analogous to
the H-equation of Chandrasekhar (which is however in one dimension
only) which is known to admit multiple solutions. It has been
shown however (Kelley 1982) that iterative methods always converge
to the only physically meaningful solution, although no
information is provided on how fast the convergence is achieved.

An important point to keep in mind is that the probability of
return from the upstream section is strongly constrained through
the condition that all particles must return to the shock, being
advected with the fluid. It follows that $$\int_{-u}^1 d\mu'
P_u(\mu_0,\mu') = 1 $$ for any value of the entrance angle
$\mu_0$. It should be stressed however that this integral constraint is
not imposed by hand. In other words, though dictated by physical
considerations, it is automatically implemented by the integral
equation for $P_u$ (Eq. \ref{eq:Pu}): it thus provides an
important numerical check of our computing accuracy.

The slow convergence of the iterative solution of Eq. \ref{mainPd}
has been partly corrected by use of the following ruse. The
conditional probability $P_d$ obeys the following integral
constraint. From Eq. \ref{eq:transport} we see that at downstream
infinity the condition $\partial f/\partial z=0$ implies that the
only physically acceptable solution for the distribution function is
$f(\mu)=constant=K$. At downstream infinity we can therefore write
\begin{equation}
f_-(\mu) (u+\mu) = K (u+\mu) = - \int_{-u_d}^1 d\mu' f_+(\mu') (u+\mu')
P_d(\mu',\mu) = - \int_{-u_d}^1 d\mu' K (u+\mu') P_d(\mu',\mu),
\end{equation}
which implies the condition
\begin{equation}
\int_{-u_d}^1 d\mu' (u_d+\mu') P_d(\mu',\mu) = - (u_d + \mu)
\label{eq:property}
\end{equation}
for any value of the exit angle $\mu$. Since the function $P_d$ depends
only on the scattering properties of the fluid, which are assumed to be
independent of $z$, Eq. \ref{eq:property} also applies to any $z$,
including $z=0$ which is the position of the shock.

This condition is of course automatically satisfied at the end of
our calculations, while it is only approximately satisfied during
the iterative solution of Eq. \ref{mainPd}. We have found out that,
by correcting the normalization of $P_d$ so that it obeys the
previous integral constraint to hold at the end of {\it every}
iteration, the convergence is much accelerated, above all in the
SPAS regime. We are unable to explain why this occurs, but
shortening of the computation times by a factor of 3, or more, have
thus been achieved.

Once the two functions $P_u$ and $P_d$ have been calculated, the slope
of the spectrum, as discussed in paper I, is given by the solution of
the equation:

\begin{equation}
(u_d+\mu) g(\mu) = \int_{-u_d}^1 d\xi Q^T (\xi,\mu) (u_d+\xi) g(\xi),
\label{eq:fred}
\end{equation}
where
\begin{equation}
Q^T(\xi,\mu) = \int_{-1}^{-u_d} d\nu P_u(\nu,\mu) P_d(\xi,\nu)
\left(\frac{1-u_{rel}\mu}{1-u_{rel}\nu}\right)^{3-s}.
\label{eq:QT}
\end{equation}
Here $u_{rel}=\frac{u-u_d}{1-u u_d}$ is the relative velocity
between the upstream and downstream fluids and $g(\mu)$ is the
angular part of the distribution function of the accelerated
particles, which contains all the information about the
anisotropy. A point to notice concerns the function $P_u$ in Eq.
(\ref{eq:QT}): all variables and functions here are evaluated in
the downstream frame, while the $P_u$ calculated though Eq.
(\ref{eq:Pu}) is in the frame comoving with the (upstream) fluid.
The $P_u$ that need to be used in Eq. (\ref{eq:QT}) is therefore
$$ P_u(\nu,\mu) = P_u(\tilde \nu , \tilde \mu) \frac{d \tilde
\mu}{d\mu}= P_u(\tilde \nu , \tilde \mu)
\left[\frac{1-u_{rel}^2}{(1-u_{rel}\mu)^2} \right]. $$ The
solution for the slope $s$ of the spectrum is found by solving Eq.
\ref{eq:fred}. In general, this equation has no solution but for
one value\footnote{The fact that the solution is unique could not
be demonstrated rigorously, but in our calculations we never found
cases of multiple solutions.} of $s$. Finding this value provides
not only the slope of the spectrum but also the angular
distribution function $g(\mu)$.

\section{Comparison with previous results}

In this section we apply the method detailed above to several cases
that have already been discussed in the literature, in order to show
that the approach can be successfully applied to a variety of cases.
More specifically, we consider the following test situations:

\begin{itemize}
\item Since our method is not based on the assumption of small pitch
angle scattering, we can demonstrate that the slope of the
spectrum of particles accelerated at Newtonian shocks is
independent of the scattering function $w(\mu,\mu')$, although the
functions $P_d$ and $P_u$ in general do depend on the choice of $w$.

\item To our knowledge, the only calculation of the functions $P_d$
and $P_u$ which applies to shocks of arbitrary velocity for the case
of large angle scattering is that of Kato \& Takahara (2001). We compare
our results for $P_d$ and $P_u$ with those presented in that paper for
a scattering function which reproduces the physical case of large angle
scattering.

\item
The slope of the spectrum of accelerated particles in the relativistic
regime for the case of large angle scattering was compared with the
results of Monte Carlo calculations carried out by Ellison, Jones
and Reynolds (1990).

\item The case of small pitch angle scattering was assumed in most
previous calculations. We test our approach in this regime in two
ways: 1) by showing that the slope of the spectrum of accelerated
particles in the SPAS approximation, as computed through our
method for several values of the shock Lorentz factor, compares
well with the results presented in (Kirk et al., 2000); 2) by
comparing our angular distribution function with that obtained by
Kirk \& Schneider (1987) for a given relativistic shock velocity.

\end{itemize}

\subsection{The Newtonian case}

The theoretical approach presented here allows us to determine the
spectrum and the anisotropy in the particle distribution for
arbitrary shock velocity and arbitrary scattering properties of
the fluid. In this section we show that we can reproduce the
results expected in the non-relativistic limit. We calculated the
spectral slope and the angular part of the distribution function
for several values of the shock velocity and shock compression
factor $r$. We also calculated the return probability from the
downstream and upstream sections. The expected return probability
in the Newtonian limit is $P_{ret}^d=1-4 u_d$, very accurately
reproduced by our calculation, for instance for $u_d=10^{-2}$
($P_{ret}^d = 0.96$) and for $u_d=2\times 10^{-2}$ ($P_{ret}^d =
0.92$). In all cases that we considered, the condition
$$\int_{-u}^1 d\mu' P_u(\mu_0,\mu') = 1 $$ is confirmed for any
value of $\mu_0$ within a part in $10^4$. As a matter of fact, we
checked that the above equality is satisfied in all cases we
studied, relativistic or otherwise.

\begin{figure}[t!]
\plottwo{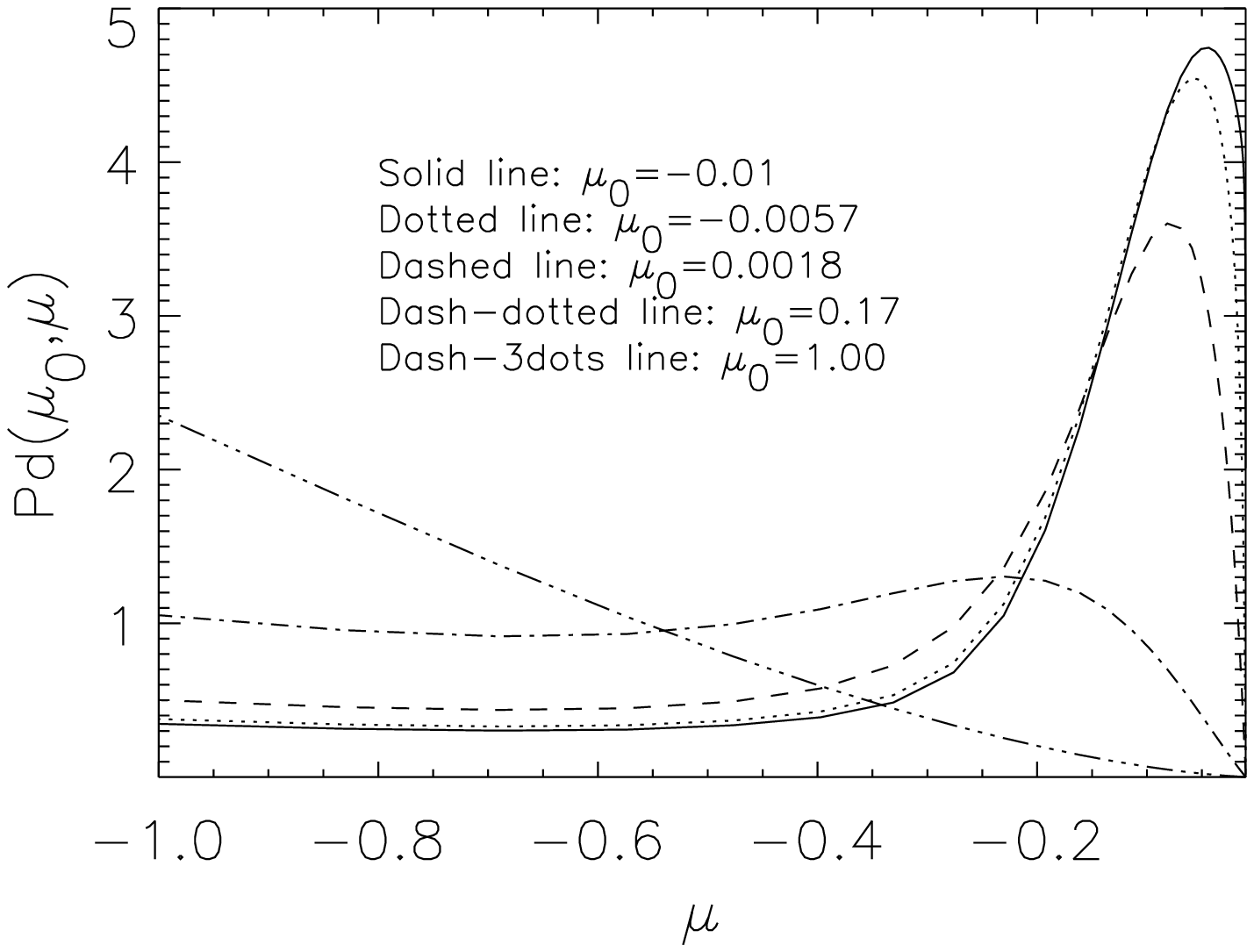}{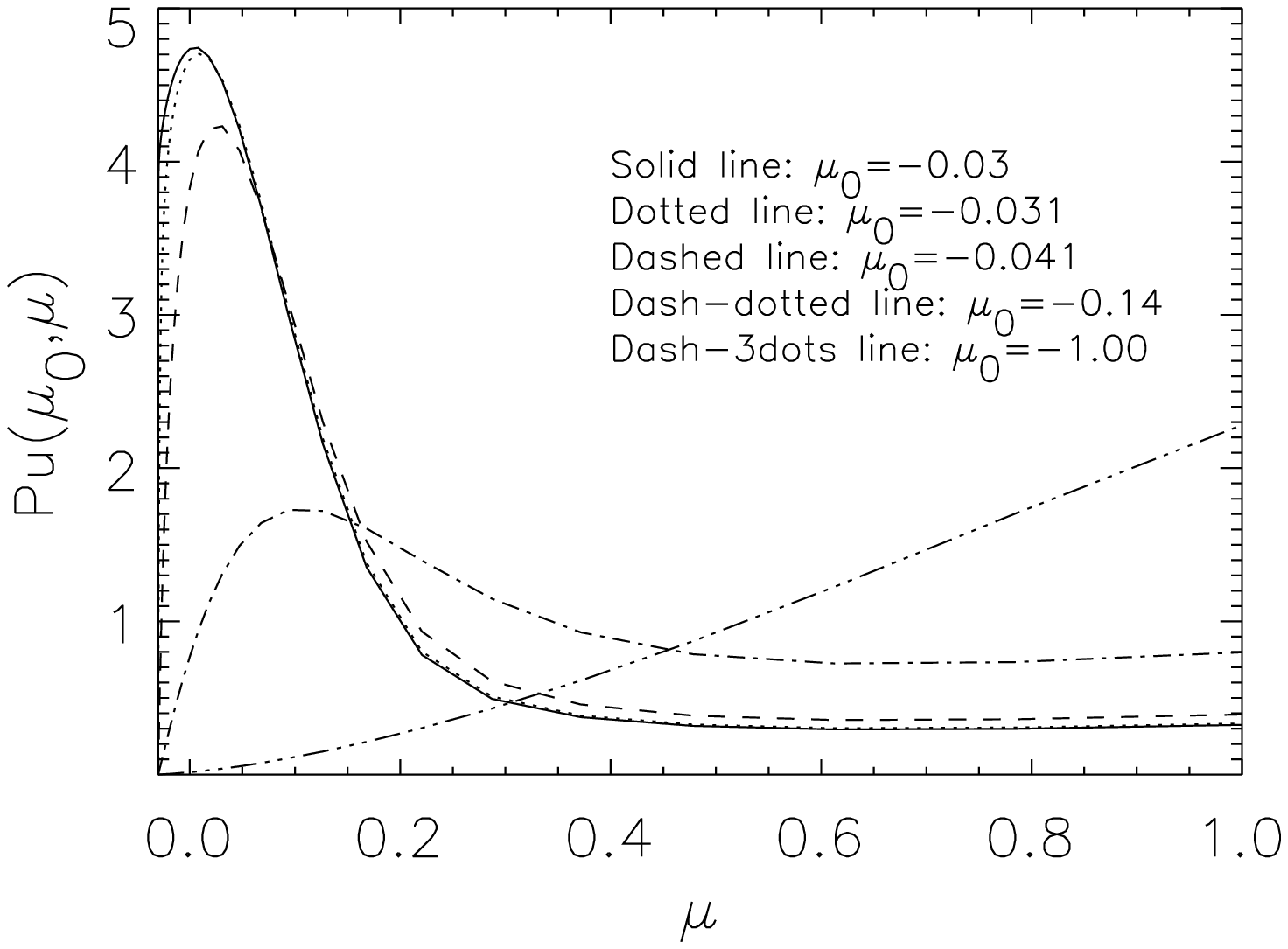}
\caption {Probability functions $P_d(\mu_0,\mu)$ and $P_u(\mu_0,\mu)$
for $u=0.03$ and $u_d=0.01$ for $\sigma=0.01$, at fixed values of $\mu_0$
(as indicated).}
\label{fig:PdPu001}
\end{figure}

\begin{figure}[t!]
\plottwo{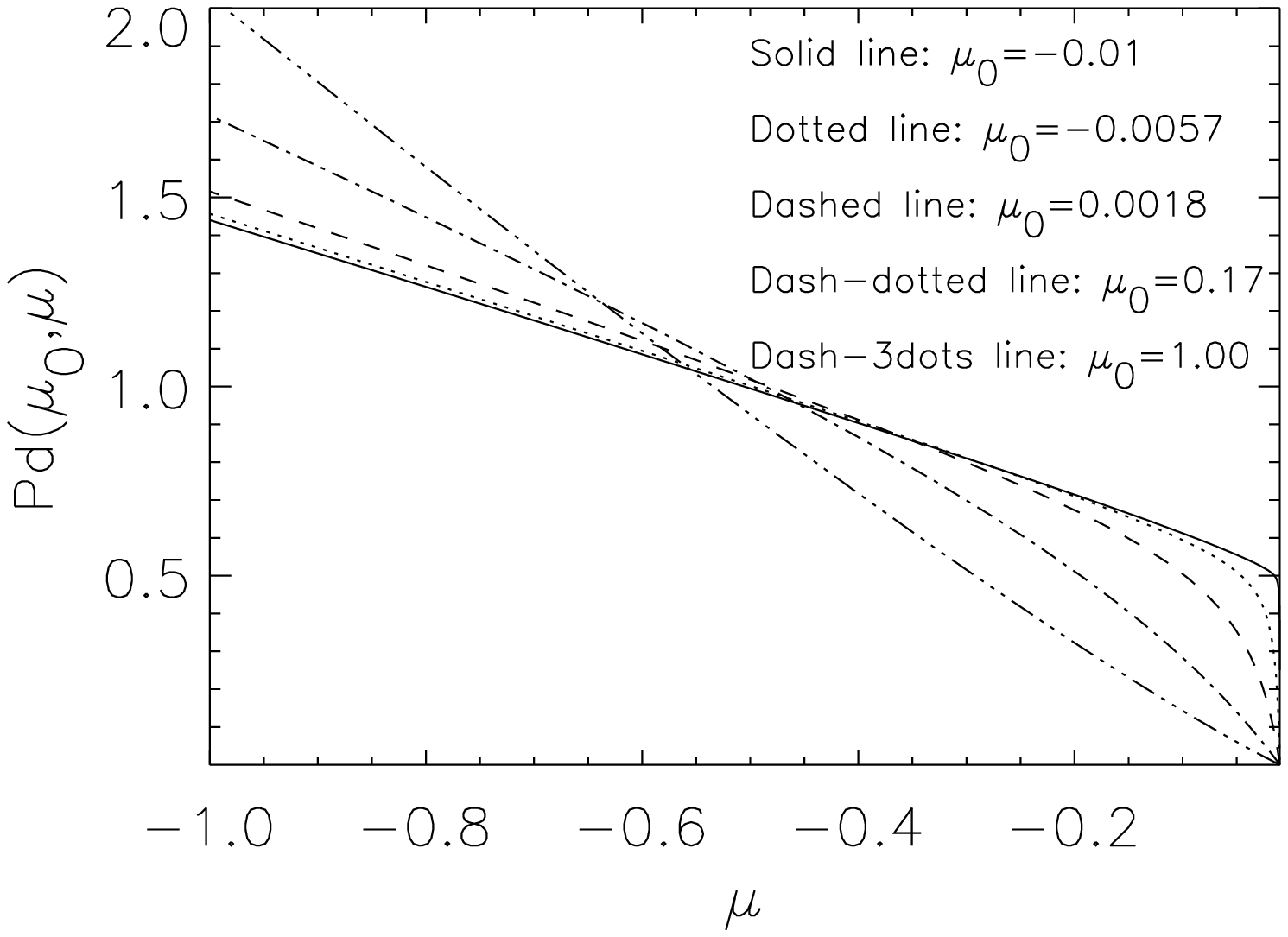}{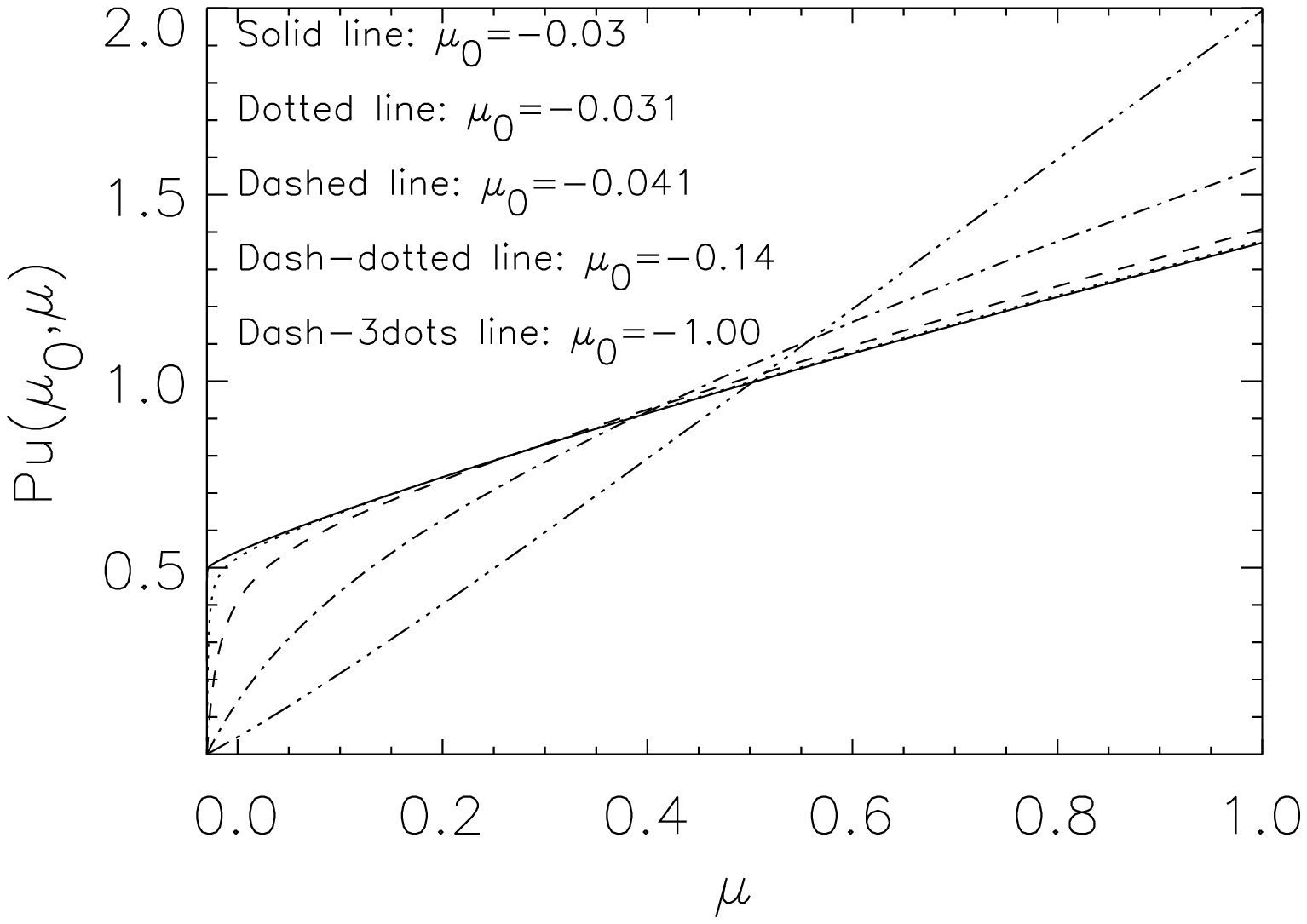}
\caption {Probability functions $P_d(\mu_0,\mu)$ and $P_u(\mu_0,\mu)$
for $u=0.03$ and $u_d=0.01$ for $\sigma=100$ (large angle scattering),
at fixed values of $\mu_0$ (as indicated).}
\label{fig:PdPu100}
\end{figure}

The slope of the spectrum of accelerated particles for $u=3\times
10^{-2}$ and $u_d=10^{-2}$ (compression factor 3) as calculated
with our approach is 4.5, in agreement with the theoretical
prediction (Bell 1978). We obtain the same slope in both cases of
small pitch angle scattering with $\sigma=0.01$ and large angle
scattering, that we simulate by choosing $\sigma=100$ (but any large
value gives the same result). The fact that in the newtonian limit
the slope of the spectrum does not depend on the scattering properties
of the fluid is also in agreement with the theoretical prediction
of Bell (1978). It is worth stressing that this universality in the
spectrum of the accelerated particles does not reflect in the
universality of the return probabilities $P_d$ and $P_u$. In Fig.
\ref{fig:PdPu001} we plot $P_d(\mu_0,\mu)$ and $P_u(\mu_0,\mu)$
as functions of $\mu$ for the values of $\mu_0$ indicated in the
figure and assuming $u=0.03,u_d=0.01$ and $\sigma=0.01$. The same
functions are plotted in Fig. \ref{fig:PdPu100} for the case of
large angle scattering ($\sigma=100$).


\subsection{The probability functions $P_d$ and $P_u$ for the relativistic
regime and large angle scattering}

As stressed above, our method is based on the introduction of the
two probability distributions $P_d$ and $P_u$, that can be calculated
by solving two integral equations, Eqs. (\ref{eq:Pd}) and (\ref{eq:Pu}).
The only other approach that we are aware of, that introduces these
two functions is that presented by Kato \& Takahara (2001), based on
the theory of random walk. We apply our calculations to reproduce Fig. 6
of their paper, obtained for a fluid speed $u=0.5$ both upstream and
downstream and in the regime of large angle scattering.

\begin{figure}[t!]
\plottwo{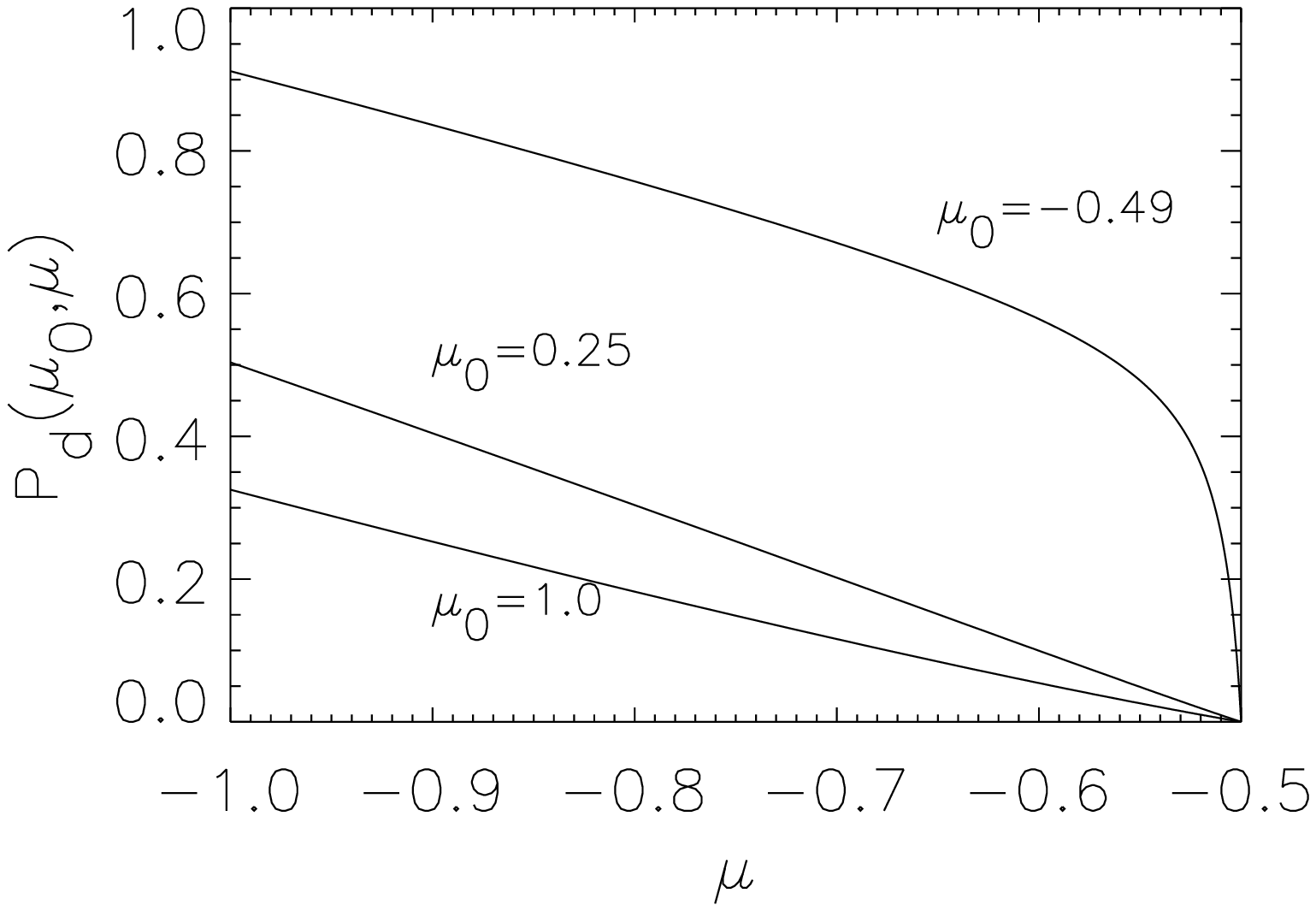}{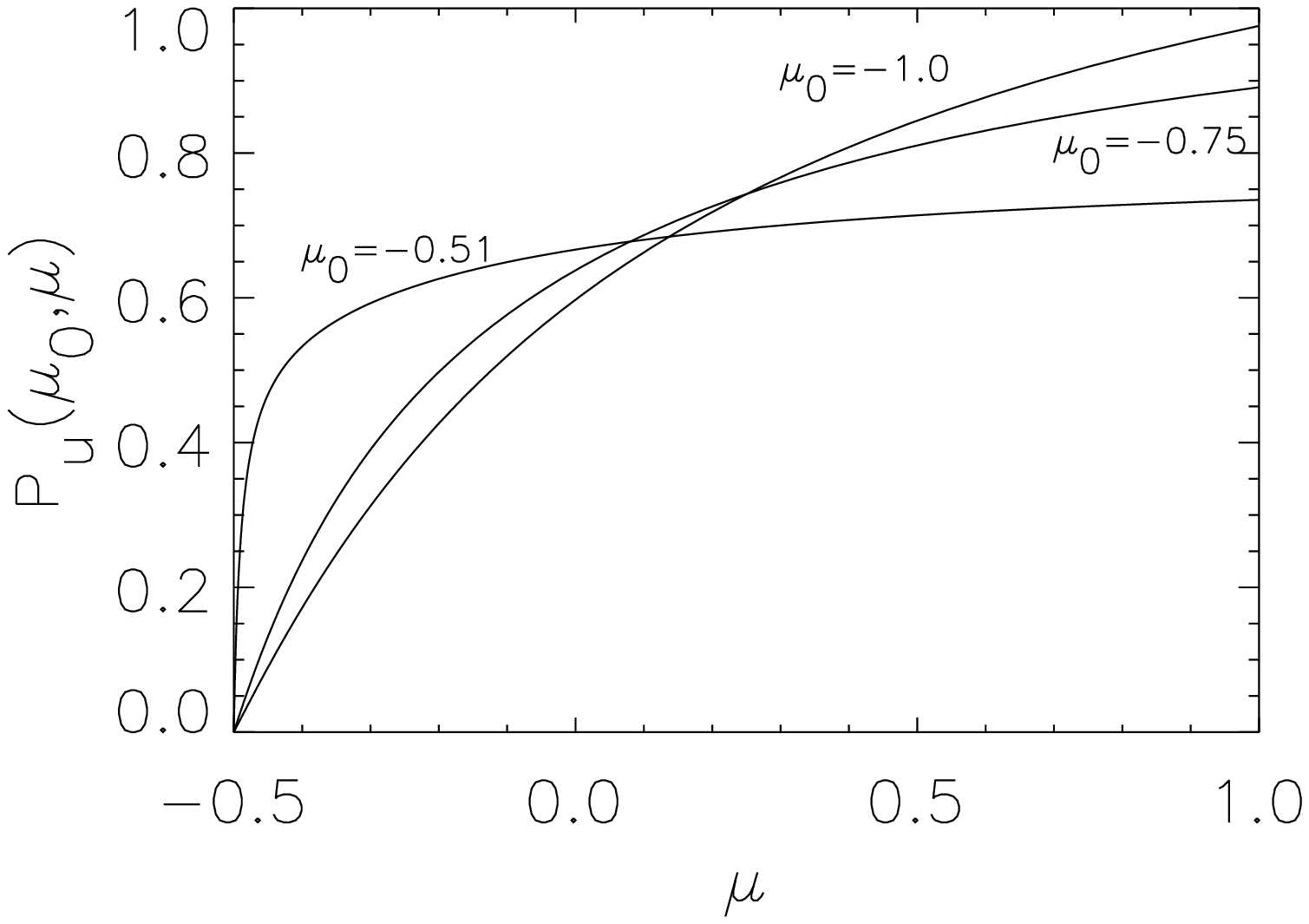}
\caption {Return probabilities $P_d(\mu_0,\mu)$ (left) and $P_u(\mu_0,\mu)$
(right) that a particle entering the downstream (upstream) region with an
angle having cosine $\mu_0$, exits it with an angle having cosine $\mu$,
for the values of $\mu_0$ as indicated.}
\label{fig:PdPu}
\end{figure}

We plot our results in Fig. \ref{fig:PdPu} for the functions $P_d$
(left panel) and $P_u$ (right panel). We carry out our calculations
for exactly the same  values of $\mu_0$ as indicated in Fig. 6 of
Kato \& Takahara (2001) and reported in our Fig. \ref{fig:PdPu} as
well. It is worths stressing that for the downstream section of the
fluid $u_d+\mu_0>0$ and $u_d+\mu<0$, while for the upstream section
these relations have opposite signs.

There is perfect agreement between our results and those of Fig. 6
of Kato and Takahara (2001), which is further evidence that our
method is perfectly able to reproduce previous results, despite the
absolute absence of any physical approximation.

\subsection{The spectrum for the case of large angle scattering}

The spectrum of the accelerated particles for the case of pure large
angle scattering was calculated by Kato \& Takahara (2001) and there
compared with the previous Monte Carlo result of Ellison et al. (1990).
The case of mixed small and large angle scattering was investigated by
Kirk \& Schneider (1988).

\begin{figure}[t!]
\plotone{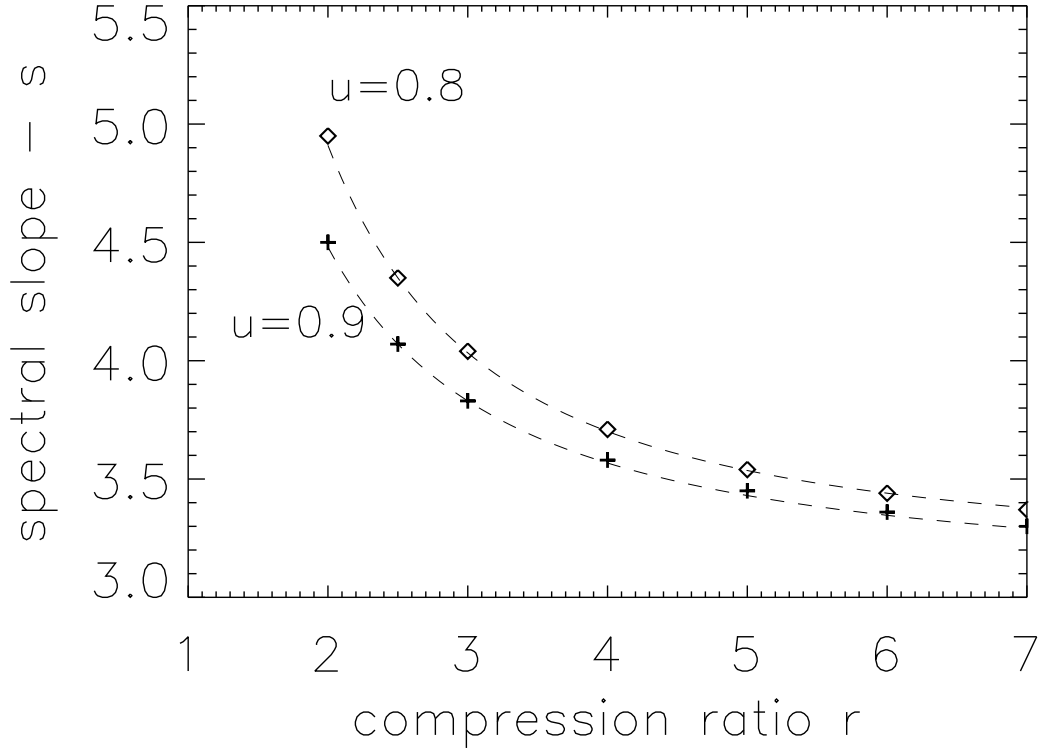}
\caption {Slope of the spectrum of accelerated particles for a shock
with speed $u=0.8$ (diamonds - upper curve) and $u=0.9$ (crosses -
lower curve) in the regime of large angle scattering, as a function
of the compression ratio. The continuous lines are the fit to the results
of Ellison et al. (1990).}
\label{fig:slopes}
\end{figure}

We apply our method to determine the spectrum of accelerated particles
for pure large angle scattering and compare our results with those
previously obtained by Kato \& Takahara (2001) and by Ellison et al. (1990),
which were slightly discrepant for relativistic shock speeds. The
spectral slopes calculated according with the procedure defined in the
previous section are plotted in Fig. \ref{fig:slopes} for shock speeds
$u=0.8$ and $u=0.9$ as functions of the compression factor $r$ at the
shock. The crosses and diamonds are the results of our
calculations, while the continuous curves represent the interpolation
provided by Ellison et al. (1990) to their Monte Carlo results.

It is remarkable that, while the probability distributions $P_d$ and
$P_u$ computed by us and by Kato and Takahara (2001) agree
perfectly, our slopes differ from theirs. We speculate that this is
due to the different method of computation. Kato and Takahara (2001)
use Peacock's recipe:
\begin{equation}
P <G>^{s-3} = 1
\end{equation}
where $<G>$ is the average energy fractional gain when a whole cycle
upstream $\rightarrow$ downstream $\rightarrow$ upstream is performed.
However, in paper I, we showed that the correct formula is
\begin{equation}
P <G^{s-3}> = 1
\end{equation}
which obviously differs from the previous one. The fact that our
computations, on the one hand, match the functions $P_d$ and $P_u$
of Kato and Takahara, but on the other one reproduce so perfectly
the slopes of Ellison, Jones and Reynolds (1990) (which use neither
our approach nor Peacock's) supports our claim that our formula is
the correct one.

\subsection{The SPAS limit}

Most calculations previously presented in the literature adopt the
assumption of SPAS, in order to reduce the transport equation (Eq.
\ref{eq:transport}) to a Fokker-Planck equation, as shown in
section \ref{sec:fp}. In the approach presented in Paper I and
here, the scattering occurs with arbitrary characteristics,
embedded in the scattering function $w(\mu,\mu')$. The case of a
$w$ independent of pitch angles is that of strong scattering,
which is expected to be achieved for large amplitude turbulence.
The case of small pitch angle scattering in the assumption of
isotropic scattering can be modeled in our approach by taking a
peaked shape for the scattering function, as explained in section
\ref{sec:numer}, with $\sigma\ll 1$.

We expect that in the limit of SPAS the result should not depend
upon the detailed form of the scattering function, provided this
is strongly peaked (see Section 3).

For a given Lorentz factor of the shock ($\gamma_{sh}$), the velocity of the
upstream fluid $u=\beta_{sh}$ is calculated. For the purpose of comparing our
results with those of Kirk et al. (2000), we adopt a Synge (1957) equation of
state for the downstream fluid and assume that the magnetic field is not
dynamically important. These are the same assumptions as in (Kirk et al. 2000).

We can therefore calculate the velocity of the downstream fluid $u_d$
from the conservation of mass, momentum and energy at the shock surface.
The values of $u$ and $u_d$ are given in Table 1. Given these velocities
and a width $\sigma$ for the scattering function, we can solve the integral
equations for $P_u$ and $P_d$ iteratively.

\begin{center}
\begin{tabular}{|c|c|c|c|} \hline
$\gamma_{sh}\beta_{sh}$ & $u$ & $u_d$ & $slope$ \\ \hline
  0.04 &   0.04  & 0.01  & 4.00 \\
    0.2 &   0.196 & 0.049 & 3.99 \\
    0.4     &   0.371 & 0.094 & 3.99 \\
    0.6 &   0.51  & 0.132 & 3.98 \\
    1.0 &   0.707 & 1.191 & 4.00 \\
    2.0 &   0.894 & 0.263 & 4.07 \\
    4.0 &   0.97  & 0.305 & 4.12 \\
    5.0 &   0.98  & 0.311 & 4.13 \\
\hline
\end{tabular}
\end{center}

Following the procedure outlined in section \ref{sec:numer}, we determine
the slope of the power law spectrum and the angular part $g(\mu)$ of the
distribution function.

\begin{figure}[t!]
\plotone{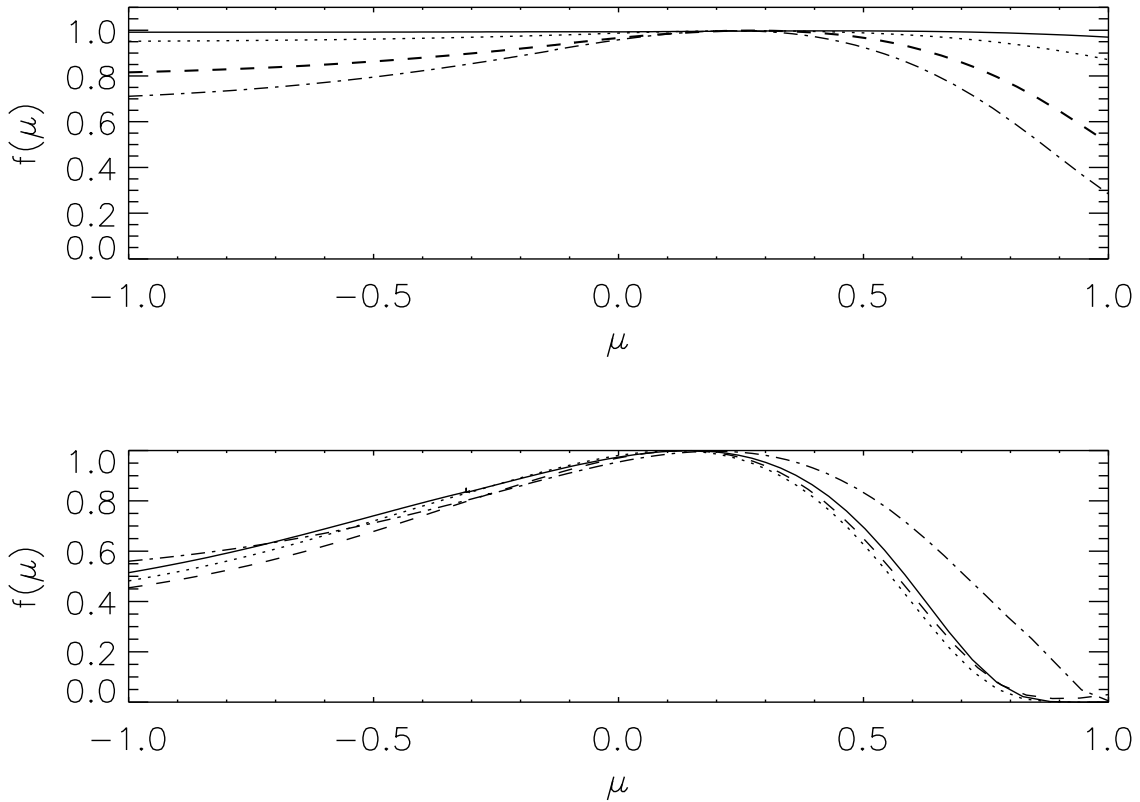}
\caption {Upper panel: DF for $\gamma_{sh}\beta_{sh}=0.04$ (solid line),
$\gamma_{sh}\beta_{sh}=0.2$ (dotted line), $\gamma_{sh}\beta_{sh}=0.4$
(dashed line) and $\gamma_{sh}\beta_{sh}=0.6$ (dash-dotted line).
Lower panel: DF for $\gamma_{sh}\beta_{sh}=1$ (dash-dotted line),
$\gamma_{sh}\beta_{sh}=2$ (dashed line), $\gamma_{sh}\beta_{sh}=4$
(dotted line) and $\gamma_{sh}\beta_{sh}=5$ (solid line).}
\label{fig:anis}
\end{figure}

The slopes predicted by our calculations for different values of
the shock velocity are shown in Table 1 for $\sigma=0.01$ and agree
well with those of Kirk et al. (2000). In Fig. \ref{fig:anis} we
also plot the distribution functions obtained for all the cases
reported in Table 1 (the upper panel refers to the first four lines
in Table 1, while the lower panel refers to the relativistic regime,
namely the last four lines in Table 1).

This figure shows the expected phenomenon of increasing anisotropy
in the distribution funtion when the shock speed increases: for
non-relativistic speeds the distribution function is almost perfectly
independent of the pitch angle $\mu$, but it becomes more and more
anisotropic in the trans-relativistic regime and in the fully relativistic
regime.

A necessary condition for the SPAS regime to be at work is that
$\sigma \ll 1/4\gamma_{sh}^2$. This implies that the SPAS assumption
requires increasingly lower values of $\sigma$ when the Lorentz factor of
the shock increases. For $\sigma = 0.01$, the maximum Lorentz factor
for which we can assume to be in the SPAS regime is $\gamma_{sh}\sim 5$,
and in fact we can already see that the slopes that we obtain for
$\gamma_{sh} \beta_{sh} = 5$ are slightly smaller than those of
Kirk et al. (2000). This effect will be discussed in detail in a
forthcoming paper.

It is worth reminding that the SPAS approximation is in fact
expected to be broken in Nature in the relativistic regime: this
is due to the fact that a FP equation describes well the
scattering only when particles scatter many times within a given
angle. In the upstream section, particles are caught up by the
shock as soon as their deflection angle with respect to the normal
to the shock is larger than $1/\gamma_{sh}$. For relativistic
shocks, this quantity becomes small and eventually comparable with
the average deflection angle per unit length, and a description by
means of a Fokker--Planck equation becomes inadequate, as remarked
many times in the literature (Kirk \etal, 2000). We stress here
that the fact that our method is not based upon the SPAS approximation,
but on the more general Eq. \ref{eq:transport}, makes it more
widely applicable then previous approaches. Its potential for a
more correct investigation of the large $\gamma$ limit will have
to be investigated elsewhere.

\begin{figure}[t!]
\plotone{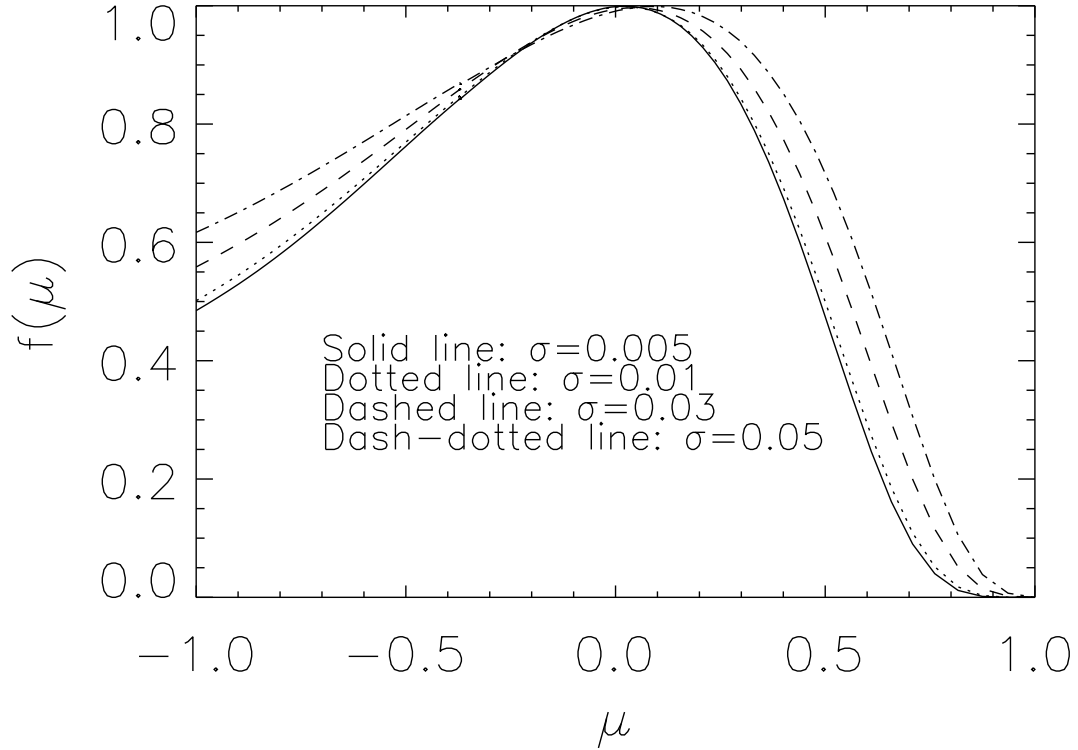}
\caption {The angular distribution function $g(\mu)$ for $u=0.9$
and for the relativistic equation of state for the downstream gas
$u u_d = 1/3$. The solid, dotted, dashed and dot-dashed lines refer
respectively to $\sigma=0.005,~0.01,~0.03$ and $0.05$.}
\label{fig:gmu}
\end{figure}

An additional evidence of how important these effects can be for the
determination of the correct distribution function of the
accelerated particles is illustrated below. We calculate here the
spectrum and the function $g(\mu)$ for $u=0.9$ and for the
relativistic equation of state for the downstream gas ($u u_d =
1/3$), as adopted by Kirk \& Schneider (1987). We carry out our
calculations for $\sigma=0.05,~0.03,~0.01$ and $0.005$. The slopes
in the four cases are very close to each other
($s=4.68,~4.69,~4.71,~4.71$) but the angular distribution functions
(normalized to be unity at the peak) appear slightly different. It
is remarkable however that when the SPAS regime is approached, the
DF converges exactly to that found by Kirk \& Schneider (1987), as
illustrated in Fig. \ref{fig:gmu}, where we plotted $g(\mu)$ for the
four cases mentioned above: compare our results with Kirk and
Schneider's Fig. 6. The slope that we obtain is also in perfect
agreement with Kirk \& Schneider (1987).

\section{Conclusions}

In Paper I, a new analytic approach to the calculation of the
spectrum of particles accelerated at a shock surface moving with
arbitrary velocity was proposed. In the present paper we described
the procedure to calculate the probability functions
$P_d(\mu_0,\mu)$ and $P_u(\mu_0,\mu)$ introduced in Paper I. These
two functions allow us to calculate the spectrum and the angular
distribution function of the accelerated particles at the shock,
for arbitrary shock velocity and for arbitrary scattering function
$w$.

We tested the approach in several situations, in order to show that
all results previously appeared in the literature could be successfully
reproduced. In particular the following tests have been performed:

\begin{itemize}
\item we confirmed the universality of the slope of the power law
spectrum of accelerated particles in the case of Newtonian shocks.
More specifically we proved that such slope is roughly independent
of the scattering properties of the medium, as described by the
scattering function $w$. It should be stressed that such
universality does not extend to the probability functions $P_d$
and $P_u$, which are instead strongly dependent upon the
scattering properties. Despite the difference between these
probability functions, their combination results in an angular
part of the distribution function of accelerated particles which
is approximately flat, as expected in the Newtonian limit
(isotropy).

\item We checked the correctness of the probability functions $P_d$
and $P_u$ in the case of large angle scattering by comparing our
results with those of Kato \& Takahara (2001). We emphasize that
these two functions contain all the physical information about the
spectrum and angular distribution of the accelerated particles.

\item We reproduce exceptionally well the results of the Monte
Carlo experiments of Ellison et al. (1990) on the slope of the
spectrum of accelerated particles as a function of the shock
compression ratio for different shock speeds, in the large angle
scattering regime, but not Kato and Takahara's (2001) results. We
ascribed this discrepancy to Kato and Takahara's use of the
(incorrect) Peacock's formula for the spectral slope.

\item We carried out calculations for the case of a scattering function
very peaked as a function of the difference between the cosines of
the entrance and exit angles. We adopted a $w$ in the form given
by our Eq. \ref{wpeaked}. The case of small $\sigma$ is expected
to describe the limit of isotropic small
pitch angle scattering, which is adopted in most literature on
shock acceleration. We showed that in this limit the transport
equation (Eq. \ref{eq:transport}) reduces to a Fokker-Planck
equation, identical to that used for instance by Kirk \& Schneider
(1987) and Kirk et al. (2000). We carried out computations for
$\sigma= 0.05-0.005$, and we compared our predicted spectral
slopes with those of Kirk et al. (2000) for different values of
the shock velocity. Our results are illustrated in Table 1 and
show that our method reproduces faily well the SPAS results.
Moreover, our approach shows the expected deviations from the
SPAS regime when the condition $\sigma\ll 1/(4\gamma_{sh}^2)$
is violated.

\item We have also compared our distribution function at the shock
with that published by Kirk and Schneider (1987), obtaining very
good agreement, and an identical slope. We also show that
the distribution function becomes increasingly different when
the strict SPAS regime is broken (Fig. \ref{fig:gmu}).

\end{itemize}

On the basis of the tests carried out to check our approach, we can
claim that this method of calculation of the spectrum and angular
distribution function of particles accelerated at shocks of {\bf
arbitrary velocity} and {\bf arbitrary} (but spatially homogeneous)
scattering properties of the fluid is fully successful in
reproducing old results, and it can now be applied to situations
which go beyond the well studied case of small pitch angle
scattering. In a forthcoming paper we will describe the potential of
this new approach to describe particle acceleration at
ultra-relativistic shocks, including the effect of a possible large
scale magnetic field.

A sincere thanks is due to the referee, whose selfless work has
considerably improved an early version of this work.

\end{document}